# The Ubiquitous Interactor – Device Independent Access to Mobile Services

*Stina Nylander, Markus Bylund and Annika Waern*
Swedish Institute of Computer Science
Box 1263
SE-164 29 Kista, SWEDEN
Tel: +46 8 633 1500
E-mail: {stina.nylander, markus.bylund, annika.waern}@sics.se

## ABSTRACT

The Ubiquitous Interactor (UBI) addresses the problems of design and development that arise around services that need to be accessed from many different devices. In UBI, the same service can present itself with different user interfaces on different devices. This is done by separating interaction between users and services from presentation. The interaction is kept the same for all devices, and different presentation information is provided for different devices. This way, tailored user interfaces for many different devices can be created without multiplying development and maintenance work. In this paper we describe the system design of UBI, the system implementation, and two services implemented for the system: a calendar service and a stockbroker service.

## INTRODUCTION

The Ubiquitous Interactor is a system addressing the problems with design and development that arise when service providers face the vast range of computing devices available on the consumer market.

Users have a wide range of devices at their disposal for accomplishing different tasks: desktop computers and laptop computers for office work, wall-sized screens for presentations in large groups, PDAs and cellular phones for mobile tasks. The range of services is equally wide: information services, shopping and entertainment. This opens for using services from different devices in different situations. Users could access for example their shopping services from a desktop computer at home and from a cellular phone on the bus. Unfortunately, this is often not possible since devices and services cannot be freely combined. Devices have different capabilities of user interaction and presentation, and most services cannot adapt their user interfaces to these differences. This means that users often have to use different versions of a service from different providers to access the same functionality. This causes problems of synchronization and compatibility.

There are two main approaches to making services accessible from multiple devices: using the same user interface on all devices, or creating a new version for each device. Both approaches have drawbacks. With the same user interface on all devices the thinnest device set the limitations of the user interface, and it is impossible to take advantage of device specific features such as scroll wheels or microphones. It is also difficult to control how user interfaces will be presented to end-users, which is important in commercial development. With a new version for each device, development and maintenance work get very cumbersome, and it is difficult to keep consistency between many different versions. Versioning allows service providers to control the presentation of user interfaces, but to the cost of more development work. We need to find new and robust methods for developing services that can adapt to different devices.

The Ubiquitous Interactor (UBI) combines the two approaches described above to create device independent services. UBI uses *interaction acts* [11] (see the design section) to describe the user-service interaction in a device independent way. This description is used by all devices to generate an appropriate user interface. The presentation of user interfaces can be controlled through *customization forms* [11] (see the design section), which contain service and device specific information of how user interfaces should be presented. This makes it possible to develop services once and for all, and tailor their user interfaces to different devices.

The rest of the paper is outlined as follows: First the background to the UBI system and some related work is discussed. Then the design decisions are described and motivated, followed by a description of the implementation of the system and services for it. Finally some conclusions are presented.

## BACKGROUND

Our interest and need for device independent services are results from our previous work with the next generation electronic services in the sView project (see below). However, the need for device independent applications is not new. During the seventies and early eighties, developers faced large differences in hardware. That time the problem disappeared when the personal computer emerged. The hardware got standardized to mouse, keyboard and desktop screen, and direct manipulation user interfaces worked similarly in different operative systems [10].

The situation that we face today is different. We are currently experiencing a paradigm shift from application based personal computing to service based ubiquitous computing [17]. In a sense, both applications and services can be seen as sets of functions and abilities that are packaged as separate units [6]. However, while applications are closely tied to individual devices, typically by a local installation procedure, services are only manifested locally on devices and made available when needed. The advance of Web based services during the nineties can be seen as the first step in this development. Instead of accessing functionality locally on single personal computers, users got used to access functionality remotely from any Internet connected PC. However, with the development of the multitude of different devices that we see today (e.g. smart phones, PDAs, and wearable computers) combined with growing requirements on mobility and ubiquity, the Web based approach is no longer enough.

For this reason, we have developed the sView system [3, 4] that provides an example of what the infrastructure for the next generation service based computing could be like. With sView, each user is provided with a personal service briefcase in which electronic services from different vendors can be stored. When accessing these services, users not only get a completely personalized usage experience, they can also benefit from the use of wide variety of different devices, continuous usage of services while switching between different devices, and network independence (completely off line use is possible).

For a long period, our only way of supporting the versatility of the range of device types in sView was to require service providers to implement many alternative user interfaces for their services. A typical end user service for example implemented a traditional GUI specified in Java Swing, an HTML and WML interface for remote access over HTTP, and an SMS interface for remote access from cellular phones. While the sView system provides support for handling transport of UI components, presentation, events, and so on, service providers still had to implement the actual user interfaces (Swing widgets, HTML/WML documents, and text messages) and interpret user actions (Java events, HTTP posts from HTML and WML forms, and text input).

This approach required great implementation and maintenance efforts of the service providers. The standard solution to the problem was no longer viable however, and alternative solutions needed to be explored. The multitude of device types we see today is not due to competition between vendors as before, but rather motivated by requirements of specialization. Different devices are designed for different purposes and thus their diverse appearance. As a result, the solution this time needs to support simple implementation and maintenance of services without loosing the uniqueness of each type of device. This is what we set out to solve with UBI.

## RELATED WORK

Much of the inspiration for the Ubiquitous Interactor comes from early attempts to achieve device independence, or in other ways simplify development work by working on a higher level than device details.

We have already mentioned that lack of hardware standards created a need of device independent applications during the seventies and the eighties. User Interface Management Systems like Mike [12] and UofA* [14] addressed this problem, together with model-based approaches like Humanoid [16]. Others proposed more partial solutions to shield developers from differences in input devices [9], or guide them in the selection of input devices and interaction techniques [7].

In current research, device independence is addressed in two different research fields, that of ubiquitous and mobile computing and that of universal access. The Ubiquitous Interactor (UBI) has its origin in the ubiquitous and mobile research, but provides solutions that can be of use in universal access too.

XWeb is a representative of work in the mobile and ubiquitous research field [13]. Inspired by the Web and Web browsers, XWeb encodes the data sent between application and client in a device independent format. Clients are responsible for the generation of user interfaces. Clients only generate user interfaces of one single type, so users get the same type of user interface to all XWeb services unless they use different clients. However, in XWeb service providers cannot control the presentation of the user interface, something that is provided in UBI.

User Interface Markup Language (UIML), is an XML compliant markup language for specification of user interfaces [1]. This description is converted to another language, for example Java or HTML. UIML differs from UBI in that its descriptions cannot take advantage of device specific features, and it only supports user-driven interaction.

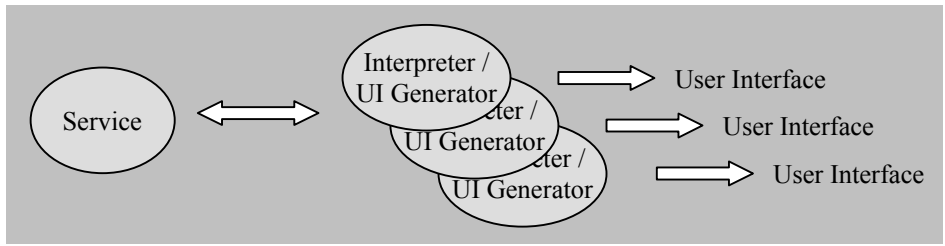

Figure 1: Services offer their interaction expressed in interaction acts, and an interpreter generates a user interface based on the interpretation. Different interpreters generate different user interfaces.

Unified User Interfaces (UUI) [15] are a representative of the universal access research community. UUI is a design and engineering framework composed by three parts: a method for design, a software architecture, and tools. The goal of UUI is to provide user interfaces tailored to different user groups and situations of use in terms of users' physical capabilities, preferences and usage context. UUI is a project with very large scope, making *all* user interfaces accessible to *all* users. This means that they take into account a large number of factors (e.g. contextual and environmental) that make the system more complex than we believe is necessary to solve the problems UBI is addressing.

## DESIGN

In the Ubiquitous Interactor (UBI), we have chosen the interaction between users and services as our level of abstraction in order to obtain units of description that are independent of device type, service type, and user interface type. Interaction is defined as *actions that services present to users, as well as performed user actions, described in a modality independent way.* Some examples of interaction according to this definition would be: making a choice from a set of alternatives, presenting information to the user, or modify existing information. Pressing a button, or speaking a command would not be examples of interaction, since they are modality specific actions. By describing the user-service interaction this way, the interaction can be kept the same regardless of device used to access a service. It is also possible to create services for an open set of devices

The interaction is expressed in interaction acts that are exchanged between services and devices. In some cases the service in question will actually be running on the device, in other cases it might be on a server. Interaction acts are interpreted on the device side and user interfaces are generated based on interaction acts and additional presentation information, see figure 1. Whether services are running locally or on a server does not affect the way services express themselves, or the way interaction acts are interpreted.

### Interaction Acts

Interaction acts are abstract units of user-service interaction that contain no information about modality or presentation. This means that they are independent of devices, services and interaction modality. Throughout this work, we assume that most kinds of interaction can be expressed using a fairly limited set of interaction acts. User-service interaction for a wide range of services can be described by combining single interaction acts and groups.

Through analysis of existing services and applications, we have defined a set of eight interaction acts that are supported in UBI: `input`, `output`, `selection`, `modification`, `create`, `destroy`, `start` and `stop`. In this definition `input` is input to the system, `output` is output to the user, `selection` is selection from a set of alternatives, and `modification` is modification of information stored in the system. `create` is creation of new objects, `destroy` is deletion of existing objects, and `start` and `stop` starts and stops the interaction with the service. All interaction acts except `output` returns user actions to services. `Output` only presents information that users cannot act upon.

During the user-service interaction, the system needs more information about the interaction acts than its type. Interaction acts need to be uniquely identifiable, so that user actions can be associated with them. Users perform actions on user interface components, and those actions need to be linked to the original interaction acts so that services can interpret them correctly. Most services will offer several interaction acts of the same type, and need a way to identify which one users acted upon. It must also be possible to define for how long a user interface component based on an interaction act should be present in the user interface and when it should be removed. Otherwise only static user interfaces can be created. It must be possible to create modal user interface components based on interaction acts, e.g. components that lock the user-service interaction until certain actions are performed by users. This way, user actions can be sequenced when needed. All interaction acts also need a way to hold default information, so that there always is something on which to base the rendering of interaction acts. Finally, it is important to be able to attach metadata to interaction acts. Metadata can for example contain domain information, or restrictions on user input that are important to the service.

In more complex user-service interaction, there is a need to group several interaction acts together, because of their related function, or the fact that they need to be presented together. An example could be the play, rewind, forward and stop functions of a CD player. The structure obtained by the grouping can be used as input when generating the user interfaces. In order to be useful, these groups should allow nesting.

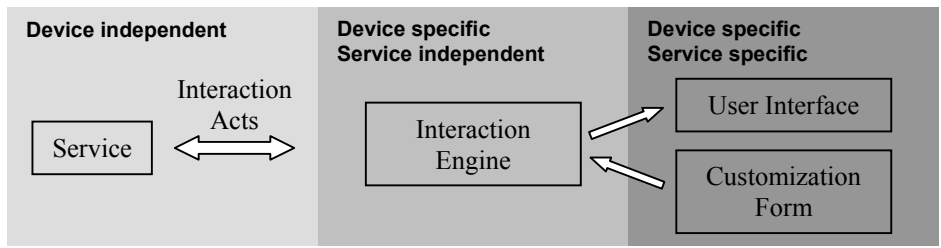

Figure 2: The three layers of specification in the Ubiquitous Interactor. Services and interaction acts are device independent, interaction engines are service independent and device or user interface specific, and customization forms and generated user interfaces are device and service specific.

The flow of interaction acts during user-service interaction is not necessarily symmetric. This means that a service for example can offer users a `modification` interaction act, but the user action performed on the `modification` interaction act can result in an interaction act of another type being returned to the service. An example is the `selection` interaction act. Some sets of alternatives include the creation of a new object or the termination of the user-service interaction. In these cases, a `selection` interaction act can return a `create` interaction act or a `stop` interaction act respectively.

**Controlling the Presentation**
To give service providers a possibility to specify how user interfaces of their services will be presented to end-users, services must be able to provide detailed presentation information. Control of presentation has proven to be an important feature of methods for developing services [5, 10], since it is used for example for branding.

In UBI, presentation information is specified separately from user-service interaction. This allows for changes and updates in the presentation information without changing the service. The main forms of presentation information are mappings and media resources. Mappings can link interaction acts to for example widgets or templates of user interface components. Media resources could be pictures or sounds that are used in the rendering of an interaction act.

It is optional to provide presentation information in UBI. If no presentation information is specified, or only partial information is provided, user interfaces are generated with default settings. However, by providing detailed information service providers can fully control how their services will be presented to end-users.

## IMPLEMENTATION
The Ubiquitous Interactor (UBI) has three main parts: the Interaction Specification Language, customization forms, and interaction engines. The Interaction Specification Language is used to encode the interaction acts sent between services and user interfaces, interaction engines interpret the encoded interaction acts and generate user interfaces, and customization forms are used to control the presentation of user interfaces. The different parts are defined at different levels of specificity, where interaction acts are device and service independent, interaction engines are device dependent, and customization forms are service and device dependent, see figure 2.

**Interaction Specification Language**
Interaction acts are encoded using the Interaction Specification Language (ISL), which is XML compliant.

Each interaction act has a unique id that is used to map performed user interactions to it. It also has a life cycle value that specifies when components based on it are available in the user interface. The life cycle can be *temporary*, *confirmed*, or *persistent*. Interface components based on temporary interaction acts are presented in the user interface for a specified time and then removed by UBI, for example a logotype shown for a few seconds when a service is starting. Interface components based on confirmed interaction acts are presented in the user interface until the user has performed a given action, for example entered required login information. Interface components based on persistent interaction acts are available in the user interface during the whole user-service interaction, or until UBI removes them. The default life cycle value is persistent. All interaction acts can be given a symbolic name, and belong to a named presentation group in a customization form. This will be discussed further in the customization form section.

Interaction acts also have a modality value that specifies if components based on them will lock other components in the user interface. The value of the modality can be *true* or *false*. If the modality value is true, the component is locking other components in the user interface until the user performs a given action, for example confirming an earlier action. The default modality value is false. All interaction acts contain a string that is used to hold default information. It is also possible to attach meta data to all interaction acts. Listing 1 shows the ISL encoding of a `selection` interaction act.

```
<selection>
  <id>235690</id>
  <life>persistent</life>
  <modal>false</modal>
  <response-number>1</response-number>
  <string>Navigation</string>
  <alternative>
    <id>98770</id>
    <string>New</string>
    <return-value>new</return-value>
  </alternative>
  <alternative>
    <id>66432</id>
    <string>Next</string>
    <return-value>next</return-value>
  </alternative>
</selection>
```

Listing 1: ISL encoding of a `selection` interaction act with id, name, life cycle, modality, and default content information. `Selection` interaction acts also contain a value for the number of alternatives that can be selected. Alternatives inherit life cycle and modality from the `selection` interaction act.

Interaction acts can be grouped using a designated tag `isl`, and groups can be nested to provide more complex user interfaces. These groups of interaction acts contain the same type of information assigned to single interaction acts: life cycle, modality, default information and meta data. Listing 2 shows the ISL encoding of a simplified example of two interaction acts grouped using the `isl` tag is shown.

```
<isl>
  <id>980796</id>
  <life>persistent</life>
  <modal>false</modal>
  <string>SICS info</string>
  <output>
    <id>235690</id>
    <life>persistent</life>
    <modal>false</modal>
    <string>SICS AB</string>
  </output>
  <output>
    <id>342564</id>
    <life>persistent</life>
    <modal>false</modal>
    <string>http://www.sics.se</string>
  </output>
</isl>
```

Listing 2: ISL encoding of two `output` interaction acts grouped using the `isl` tag.

The ISL code sent from services to interaction engines contains all information about the interaction acts: id, name, group, life cycle, modality, and metadata. A large part of this information is only useful for the interaction engine during generation of user interfaces. There is no point in sending information concerning user-service interaction handling back to the service. Thus, when users perform actions, only the relevant parts of interaction acts are sent back to the service. This includes the id for all interaction acts and for those interaction acts that imply user data input it also includes the data, for example the value of the selected alternative in `selection` interaction acts, the parameters of `create` interaction acts, or other input data. Two different DTDs have been created for this purpose, one for encoding interaction acts sent from services to interaction engines, and one for encoding interaction acts sent from interaction engines to services. The DTDs are available at http://www.sics.se/~stny/UIB/DTDs/dtd.html.

**Customization Forms Implementation**

Customization forms contain device and service specific information about how the user interface of a given service should be presented. Information can be specified on three different levels: group level, type level or name level. Information on group level affects all interaction acts of a group, and can be used to provide a look and feel for whole services or parts of services. Information at interaction act type level provides rendering information for all interaction acts of the given type; and information on name level provides rendering information about all interaction acts with the given symbolic name. The levels can also be combined, for example creating specifications for interaction acts in a given group of a given type, or in a given group with a given name.

The Interaction Specification Language provides means for creating the different mappings. Each interaction act or group of interaction acts can be given an optional symbolic name that is used in mappings where the name level is involved. This means that each interaction act with a certain name is presented using the information mapped to the name. Interaction acts or groups of interaction acts can also belong to a named group in a customization form. All interaction acts that belong to a group are presented using the information associated with the group (and possibly with additional information associated with their name or type).

```
<selection>
  <id>235690</id>
  <name>nextSelect</name>
  <group>calendar</group>
  <life>persistent</life>
  <modal>false</modal>
  <response-number>1</response-number>
  <string>Navigation</string>
  <alternative>
    ...
  </alternative>
  <alternative>
    ...
  </alternative>
</selection>
```

Listing 3: ISL encoding of an `output` interaction act with a symbolic name, and that belongs to a customization form group.

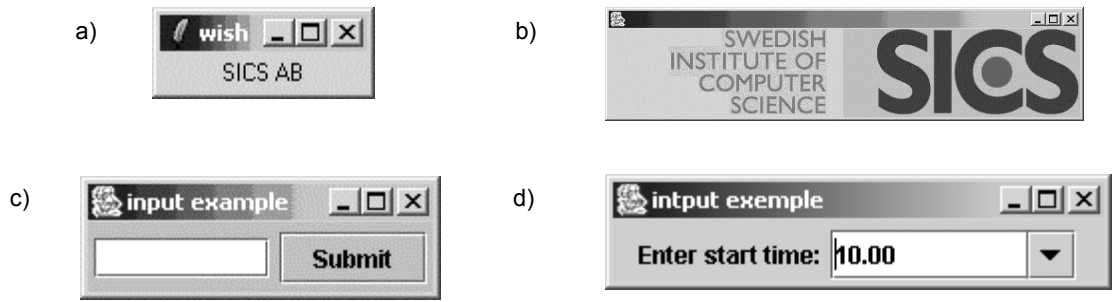

Figure 3: Rendering examples of an output and an input interaction act. Picture a and b are renderings of an output interaction act, and picture c and d are renderings of an input interaction act. is a Tcl/Tk label using the default information of the interaction act, while picture b is a Java Swing label displaying an image specified in the customization form. Picture c is a Java Swing text field with a button to submit entered text, while picture d is a Java Swing label and an editable combobox for choosing or entering time expressions.

Listing 3 shows a shortened encoding of the selection interaction act from listing 1 with a symbolic name, and as a member of the customization form group.

Customization forms are structured, and can be arranged in hierarchies. This allows for inheriting and overriding information between customization forms. A basic form can be used to provide a look and feel for a family of services, with different service specific forms adding or overriding parts of the basic specifications to create service specific user interfaces.

Customization forms are encoded in XML and a DTD can be found at http://www.sics.se/~stny/UBI/DTDs/dtd.html. An entry in a customization form can be either a *directive* or a *resource*. Directives are used for mappings to widgets or other user interface components and resources are used to associate media resources to interface components. Both directive mapping and resource association can be made on all three levels, group, type and name. Listing 4 shows an example of a directive mapping based on the type of the interaction act, in this case `output`.

```
<element name="output">
  <directive>
    <data>
      se.sics.ubi.swing.OutputLabel
    </data>
  </directive>
</element>
```

Listing 4: A mapping on type level for an output interaction act.

A customization form does not need to be complete. Interaction acts that have no presentation information specified in the form are rendered with defaults.

### An Example

To illustrate the user-service interaction in more detail we will examine an example. The `selection` interaction act in listing 2 has a name that can be used in mappings in customization forms. Listing 5 shows a sample mapping on name level from a customization form.

```
<id name="nextSelect">
  <directive>
    <data>
      se.sics.ubi.swing.SelectButton
    </data>
  </directive>
<id>
```

Listing 5: A mapping on name level in a customization form.

This mapping instructs the interaction engine to use a certain widget when presenting the interaction act. The generated presentation could look like figure 4.

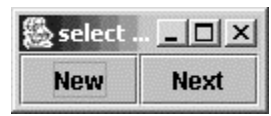

Figure 4: An example rendering of the `select` interaction act in listing 2.

We can imagine that this interaction act is used to browse a list of items using two different operations: new and next. If a user pressed the New button, a create interaction act would be returned to the service. Listing 6 shows the ISL encoding of a `create` interaction act that is to be returned to a service.

```
<create>
  <id>98770</id>
</create>
```

Listing 6: A `create` interaction act with no parameters returned to a service.

The service would interpret the interaction act, create the new object, and update the user interface if necessary.

### Interaction Engines Implementation

Interaction engines interpret interaction acts and generate suitable user interfaces of a given type for services on a given device or family of devices. Interaction engines also encode performed user actions as interaction acts and send them back to services. Examples of interaction engines are an engine for Web user interfaces on desktop and laptop computers, and an engine for Java Swing GUIs on handheld computers.

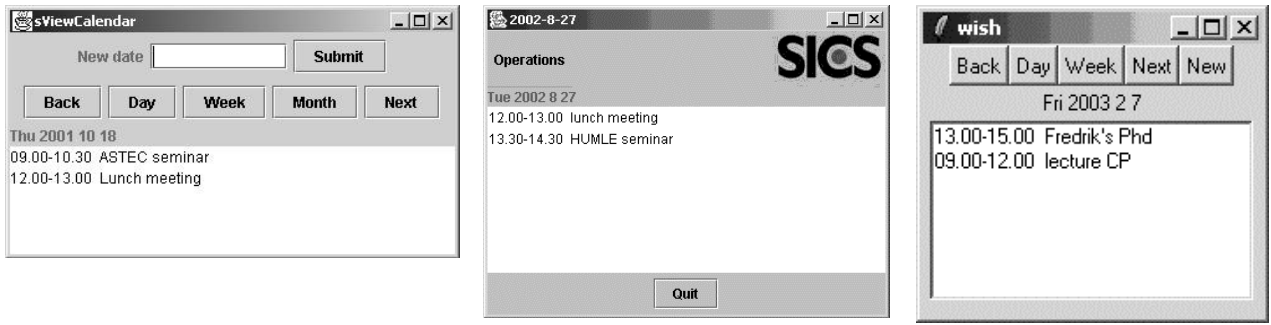

Figure 5: Three different user interfaces to the calendar service generated from the same interaction acts. The two to the left are generated by the Java Swing interaction engine using two different customization forms. The one to the right is generated by the Tcl/Tk interaction engine.

During user-service interaction, interaction engines parse interaction acts sent by services, and generate user interfaces by creating presentations of each interaction act. If specific presentations, or media resources, are specified for an interaction act in the customization form of a service, that presentation is used. Otherwise, interaction engines have defaults for each type of interaction act. For example, an output could be rendered as a label, or output generated from its default information, while an input could be rendered as a text field or a standard speech prompt. If there is a customization form, its information always takes precedence, but forms do not need to specify renderings for all interaction acts of a service. Defaults are used for the missing parts. Figure 3 shows presentations of an output and an input interaction act. The output interaction act is presented as a Tcl/Tk label showing the default information of the interaction act, and as a Java Swing label displaying an image specified in the customization form (picture a and b). An alternative presentation could be generated speech saying "SICS AB". The input interaction act is presented using Java Swing as a text field with a submit button, and an editable combo box with a text label (picture c and d).

We have implemented interaction engines for Java Swing, HTML, and Tcl/Tk user interfaces. All three interaction engines can generate user interfaces for desktop computers. The default renderings of the Tcl/Tk interaction engine are designed to create user interfaces suitable for PDAs.

*Java Swing Interaction Engine* The Java Swing interaction engine creates Java Swing widgets based on interaction acts and customization forms. Mappings are made between single interaction acts and widgets, as well as between groups of interaction acts and widgets. Mappings can be made to single widgets (e.g. a button) or to complex ones (e.g. panels with many widgets in). The Swing interaction engine can make use of both the specified lifecycle and modality of interaction acts. Interaction acts with confirmed life cycle can be rendered in a dialog window, and if the interaction acts are modal that dialog window can be made modal.

*HTML Interaction Engine* The HTML interaction engine translates between interaction acts and HTML code and user feedback is handled with HTML Forms. The nature of HTML user interfaces does not support all features of interaction acts. Since HTML user interfaces are user-driven and non-modal, the different life cycle and modality values of interaction acts are not supported.

*Tcl/Tk Interaction Engine* The Tcl/Tk interaction engine generates Tcl/Tk code based on interaction acts and customization forms to produce graphical user interfaces for PDAs. The code is executed by a small tcl client running on the device. User actions are encoded in an internal format that is converted to interaction acts by the interaction engine and sent back to services. Mappings in customization forms are made between interaction acts, and chunks of Tcl/Tk code. The Tcl/Tk interaction engine is currently not using the life cycle or modality information of the interaction acts. The Tcl/Tk interaction engine is not running on the PDA. Instead, it is running on the same machine as the service, and the generated Tcl/Tk code is sent to the device over a socket connection. Our test machine has been a Compaq Ipaq 3850 with a Tcl/Tk version for Windows CE available from http://www.rainer-keuchel.de/wince/tcltk-ce.html.

## SERVICES

We will present two different services to illustrate how the Ubiquitous Interactor (UBI) works, a calendar service and a stockbroker service.

### Calendar Service

The calendar service was the first service created for UBI. It provides a good example of a service that it is useful to access from different devices. Calendar information may often be entered from a desktop computer at work or at home, but mobile access is needed to consult the information on the way to a meeting or in the car on the way home. Sometimes appointments are set up out of office (in meeting rooms or restaurants) and it is practical to be able to enter that information immediately and not wait to get back to the office.

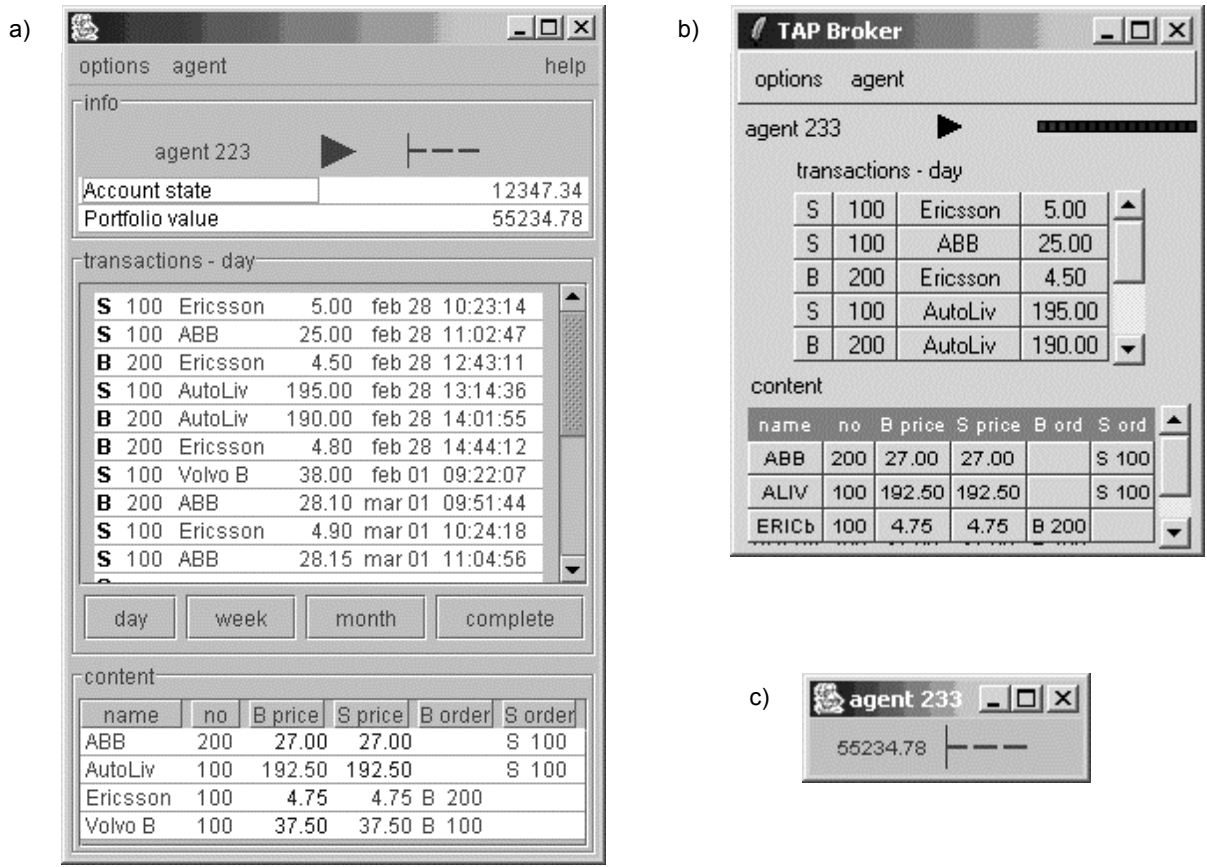

Figure 6: Three different examples of user interfaces to the TAP Broker service. Picture a shows a Java Swing user interface for desktop or laptop computers, picture b a Tcl/Tk user interface for PDA, and picture c a Java Swing user interface for very small devices (for example Java enabled cellular phones). All three user interfaces are based on the same interaction acts.

The calendar service supports basic calendar operations as entering, edit and delete information, navigate the information, and display different views of the information. The service is accessible from three types of user interfaces: Java Swing and HTML user interfaces for desktop computers, and a Tcl/Tk user interface for handheld computer. Two different customization forms have been created for Java Swing, and one each for Tcl/Tk and HTML. An example of different presentations could be a `selection` interaction act presented as a panel with five buttons (back, day view, week view, month view, next) in one of the Swing UIs, as a pull-down menu in the other, and with only four buttons in the PDA UI (a decision on customization form level not to present a month view on the PDA) (see figure 5). These different presentations are created from the same interaction act, combined with different presentation information.

**Stockbroker Service**

The stockbroker service TAP Broker has been developed as a part of a project at SICS that works with autonomous agents that trade stocks on the behalf of users [8]. Autonomous agents trade stocks on the behalf of users. Each agent is trading according to a built in strategy (for example *buy low, sell high*, or *buy and hold* [2]), and users can have one or more agents trading for them. Since agents are autonomous, users cannot control them other than contacting the agent trade server master and ask to get the agent shut down. Our service provides users with feedback on how their agents are performing so that they know when to change agent, or shut them down.

The TAP Broker service provides agent owners with feedback on the agent's actions: order handling of the agent (placing and cancelling orders), and transactions performed by the agent (buying or selling stocks). It also provides information about the agent's state: the account state (the amount of money it can invest), status (running or paused), activity level (number of transactions per hour), portfolio content, and the current value of the portfolio. However, it does not provide any means to configure or control the agent. The agents are created to work autonomously and cannot be manipulated from outside for security reasons.

We have implemented customization forms for Java Swing,

Tcl/Tk and HTML (see figure 6 for example pictures). For Java Swing, two quite different customization forms have been developed: one that generates a user interface appropriate for desktop screens, and one that generates a user interface for very small devices like java enabled cellular phones. Since the screen size and presentation capabilities of desktop computers, PDAs and cellular phones are very different, user interfaces for the smaller devices only present parts of the available information.

*The Java Swing Desktop User Interface*  The user interface generated from the desktop customization form provides updated information about all the actions of the trading agent, and about the account and the portfolio. The state of the agent, and its level of activity is also shown, see figure 6, picture a. It can provide a history of transactions in different views (current day, latest week, latest month and complete history) in a new window. Users can also switch between agents if they own more than one. This user interface is not intended to cover the whole screen, but to be present on the screen while users attend to other tasks.

*The Java Swing Small Device User Interface*  The user interface generated from the small device customization form shows considerably less information that the desktop user interface. To minimize the window, only the value of the portfolio, the state of the agent, and the activity level is shown. The value of the portfolio is color coded, red for downward trend and blue for upward trend, see figure 6, picture c. As for the desktop user interface, the purpose of this user interface is not to use small devices maximal screen resources but to be present and still leave room for other interaction.

*The HTML User Interface*  The HTML user interface displays all available information about the current agent: transactions, orders, account state, and portfolio content and value. It also provides information about the state and the activity level of the agent. As in the Java Swing desktop user interface, transaction history can be presented in different views (latest day, latest week, latest month and complete history). Due to the nature of HTML user interfaces, the information cannot be updated through system push. Updates will be made upon user actions. This means that temporary life cycle of interaction acts is not supported.

*The Tcl/Tk User Interface*  The Tcl/Tk user interface is designed for PDA use, and thus a smaller screen. To adapt to this, the Tcl/TK user interface does not show the account state and the portfolio value. A smaller number of transactions are shown, and the buttons for choosing different transaction history views are rendered as menu alternatives in the option menu, see figure 6, picture b.

**FUTURE WORK**
In the TAP Broker service, there is a great difference in the amount of information presented in different user interfaces. However, all interaction engines get the same interaction acts, thus the same amount of information, to base their user interfaces on. Thus, in those cases when the interaction engine is running on the device, and the service is running remotely, lots of superfluous interactions are sent to an interaction engine. This could be a problem when network capacity is limited. We will look at ways of server side filtering for those cases to avoid sending interaction acts that will not be used in the generation process.

Adaptation of user interfaces to device features and capabilities need to be combined with service personalization. User preferences must affect the way services present themselves. Preferences can be collected by letting users set up profiles, or by monitoring user interaction. We believe that customization forms can be used for personalization in UBI. User preferences could be stored in separate customization forms that interaction engines combined with other presentation information when generating user interfaces. Customization forms for personalization would be device and service specific just as the forms created by service providers.

We will also investigate how to handle dynamic resources in UBI. Services that use lots of dynamic media resources, e.g. a service for browsing a video database, might need an extension of our customization form approach to work efficiently for different modalities. One solution could be to handle the choice of media type outside the customization form.

**CONCLUSION**
We have presented the Ubiquitous Interactor (UBI), a system for development of device independent mobile services. In UBI, user-service interaction is described in a modality and device independent way using interaction acts. The description is combined with device and service specific presentation information in customization forms to generate tailored user interfaces. This allows service providers to develop services once and for all, and still provide tailored user interfaces to different services by creating different customization forms. Development and maintenance work is simplified since only one version of each service need to be developed. New customization forms can be created at any point, thus services can be developed for an open set of devices.

**ACKNOWLEDGMENTS**
This work has been funded by the Swedish Agency for Innovation Systems (www.vinnova.se). Thanks to the members of the HUMLE laboratory, in particular Anna Sandin for help with the implementation of the HTML interaction engine.

**REFERENCES**
1.    Abrams, M., Phanouriou, C., Batongbacal, A.L., Williams, S.M. and Shuster, J.E. UIML - an appliance-independent    XML    user    interface language. *Computer Networks*, *31*. 1695-1708.

2.  Boman, M., Johansson, S. and Lybäck, D. Parrondo Strategies for Artificial Traders. in Bradshaw ed. *Intelligent Agent Technology*, 2001, 150-159.

3.  Bylund, M. Personal Service Environments - Openness and User Control in User-Service Interaction *Department of Information Technology*, Uppsala University, Uppsala, 2001.

4.  Bylund, M. and Espinoza, F., sView - Personal Service Interaction. in *5th International Conference on The Practical Applications of Intelligent Agents and Multi-Agent Technology*, (2000).

5.  Esler, M., Hightower, J., Anderson, T. and Borriello, G., Next Century Challenges: Data-Centric Networking for Invisible Computing. The Portolano Project at the University of Washington. in *The Fifth ACM International Conference on Mobile Computing and Networking, MobiCom 1999*, (1999).

6.  Espinoza, F. Individual Service Provisioning *Department of Computer and Systems Science*, Stockholm University/Royal Institute of Technology, Stockholm, 2003.

7.  Foley, J.D., Wallace, V.L. and Chan, P. The Human Factors of Computer Graphics Interaction Techniques. *IEEE Computer Graphics and Applications*, *4* (6). 13-48.

8.  Lybäck, D. and Boman, M. Agent trade servers in financial exchange systems. *ACM Transactions on*

9.  *Internet Technology* (In press.).

9.  Myers, B.A. A New Model for Handling Input. *ACM Transactions on Information Systems*, *8* (3). 289-320.

10. Myers, B.A., Hudson, S.E. and Pausch, R. Past, Present and Future of User Interface Software Tools. *ACM Transactions on Computer-Human Interaction*, *7* (1). 3-28.

11. Nylander, S. and Bylund, M., Providing Device Independence to Mobile Service. in *7th ERCIM Workshop User Interfaces for All*, (2002).

12. Olsen, D.J. MIKE: The Menu Interaction Kontrol Environment. *ACM Transactions on Graphics*, *5* (4). 318-344.

13. Olsen, D.J., Jefferies, S., Nielsen, T., Moyes, W. and Fredrickson, P., Cross-modal Interaction using XWeb. in *UIST 2000*, (2000).

14. Singh, G. and Greene, M., A high-level user interface management system. in *Conference on Human Factors and Computing Systems, CHI 89*, (1989), 133-138.

15. Stephanidis, C. The Concept of Unified User Interfaces. in Stephanidis, C. ed. *User Interfaces for All - Concepts, Methods, and Tools*, Lawrence Erlbaum Associates, 2001, 371-388.

16. Szekely, P., Luo, P. and Neches, R., Beyond Interface Builders: Model-Based Interface Tools. in *INTERCHI'93*, (1993), 383-390.

17. Weiser, M. The Computer for the 21st Century. *Scientific American* (September 1991).